\documentclass[12pt]{article}
\usepackage{epsfig}
\usepackage{psfrag}
\usepackage{latexsym}
\usepackage{amssymb}
\usepackage{amsmath}
\usepackage{mathrsfs}

\def\be{\begin{equation}}
\def\ee{\end{equation}}
\def\bc{\begin{center}}
\def\ec{\end{center}}
\def\bea{\begin{eqnarray}}
\def\eea{\end{eqnarray}}

\newcommand{\ba}{\begin{array}{c}}
\newcommand{\bad}{\begin{array}{ccc}}
\newcommand{\ea}{\end{array}}

\def\nn{\nonumber}
\def\dd{\displaystyle}

\begin{document}

\begin{titlepage}
\hfill{RM3-TH/11-6}

\vskip 2.5cm
\begin{center}
 {\Large\bf  Bimaximal mixing and large $\theta_{13}$ in a SUSY SU(5) model based on $S_4$}
 \end{center}
\vskip 0.2  cm
\vskip 0.5  cm
 \begin{center}
 {\large Davide Meloni }~\footnote{e-mail address: davide.meloni@fis.uniroma3.it}
 \\
 \vskip .2cm {\it Dipartimento di Fisica "E. Amaldi"}
 \\
 {\it Universit\'a degli Studi Roma Tre, Via della Vasca Navale 84, 00146 Roma, Italy}
 \\
 \end{center}
\vskip 0.7cm

 \begin{abstract}
The recent analyses of the world neutrino data, including the T2K and MINOS results, point toward a statistically significant deviation of $\theta_{13}$ from zero.
In this paper we present a SUSY SU(5) model based on the discrete $S_4$ group which predicts a 
large $\theta_{13}\sim {\cal O}(\lambda_C)$, $\lambda_C$ being the Cabibbo angle. The other mixing angles in the neutrino sector are all compatible with current experimental data. 
In the quark sector, the entries of the CKM mixing matrix as well as the mass hierarchies in both up 
and down quark sectors are well reproduced and only a small enhancement is needed to reproduce $\lambda_C$. 

 \end{abstract}
 \end{titlepage}
\setcounter{footnote}{0}
\vskip2truecm

\section{Introduction and description of the model}
\label{description}
The T2K experiment has recently observed six events which, after all selection criteria at the far detector, are a strong indication of $\nu_\mu \to \nu_e$ 
flavour transition \cite{T2K}. In a three flavour scenario with $|\Delta m _{23}^2| = 2.4 \times 10 ^{-3}$ eV$^2$ and maximal atmospheric mixing, these data are consistent 
with a non-vanishing $\theta_{13}$ at 2.5$\sigma$, with $0.03(0.04)<\sin^2 2\theta_{13}<0.28(0.34)$ for normal (inverted) hierarchy and best fit value 
at 0.11 (0.14). 
This result goes in the same direction of other, statistically less significant, analyses where a not-so-small
reactor angle emerged from global fits to the available neutrino data \cite{Schwetz:2011,Fogli:Indication,GonzalezGarcia:2010er}.

Hints for $\theta_{13}>0$ have been confirmed in \cite{new}, where an analysis of global neutrino data, including the latest
T2K and MINOS results \cite{minos}, provides 
\begin{equation}
\sin^2\theta_{13}=\left\{ 
\begin{array}{l}
0.021\pm 0.007\, ,\ \mathrm{old\ reactor\ fluxes}\\[1mm]
0.025\pm 0.007\, ,\ \mathrm{new\ reactor\ fluxes}
\end{array}
\right. \ (1\sigma)\ , 
\end{equation}
which corresponds to  a $> 3\sigma$ evidence for a non-vanishing reactor angle, for both old and new reactor neutrino fluxes \cite{Mueller:2011nm}.
A large value of $\theta_{13}$ seems
to disfavor the picture where the tri-bimaximal (TBM) mixing pattern
\cite{hps},($\tan^2 \theta_{23} = 1, \tan^2 \theta_{12} = \frac{1}{2}$ and $\sin \theta_{13} =0$), is a good first order description of the data.
In fact, models based on $A_4$, $S_4$, $T_7$, $T^{\prime}$ and  so on (see \cite{Altarelli:2010gt} for a review) have the common feature 
that, at the next level of approximation, all the three mixing angles receive corrections of 
the same order of magnitude, which is basically fixed by the experimentally allowed departures of $\theta_{12}$ from its TBM value, at the level
of $\mathcal{O}(\lambda_C^2)$, with $\lambda_C$ being the Cabibbo angle. As a consequence, $\theta_{13}$ (and also the deviation of $\theta_{23}$ from the 
maximal value) is expected to also be at most of $\mathcal{O}(\lambda_C^2)$, which is now only marginally allowed (but see \cite{Shimizu:2011xg} for some examples of 
large $\theta_{13}$ in the context of TBM). 
As it has been pointed out in \cite{Altarelli:2009gn}, a large value of the reactor 
angle can be achieved if the Bimaximal (BM) mixing option, $\tan^2 \theta_{23} = 1, \tan^2 \theta_{12} = 1, \sin \theta_{13} =0$, is the correct first order approximation 
to describe the neutrino mixings; corrections from the charged lepton sector 
must be large enough, of $\mathcal{O}(\lambda_C)$, to move $\theta_{13}$ toward large values and to reconcile the value of $\theta_{12}$ with the experiments but much smaller
in the atmospheric sector in order not to destroy the agreement with maximal mixing. This pattern of subleading corrections is in fact realized in \cite{Altarelli:2009gn}, giving:
\bea
\sin^2\theta_{12}&\sim&\frac{1}{2} + {\cal O}(\lambda_C) \qquad \sin^2\theta_{23}\sim\frac{1}{2}+ {\cal O}(\lambda_C^2) \qquad
\sin\theta_{13}\sim {\cal O}(\lambda_C) \,.
\label{sinNLO}
\eea
In this paper we consider a GUT extension of such a model, based on the SU(5) gauge group, with the aim of 
extending the $S_4$ symmetry to describe the quark sector and maintaining at the same time the relations in eq.(\ref{sinNLO}).
Earlier attempts to describe quarks and leptons in the context of SU(5)$\times S_4$ can be found in \cite{su5s4} where, however, the main goal was to 
reproduce the TBM. Similar tentatives in SO(10) and Pati-Salam have been discussed in \cite{Patel:2010hr} and
\cite{Toorop:2010yh}, respectively.
The model proposed here is the first attempt to get the BM pattern in a SU(5) context based on $S_4$.
In order to formulate a realistic GUT model, we work in a supersimmetric scenario in 4+1 dimensions where problems related 
to the breaking of a grand unified symmetry (like the doublet-triplet splitting problem and the proton decay) can be efficiently solved  \cite{5DSU5}.
In the simplest setting, the fifth dimension is compactified on a circle $S^1$ of radius $R$ in such a way that
the gauge fields, living in the whole 5D space-time, are assumed to be periodic along the extra dimension only up to
a discrete parity transformation $\Omega$ under which  the gauge fields of the SU(3)$\times$SU(2)$\times$U(1) subgroup are periodic 
and possess a zero mode while those of the coset SU(5)/SU(3)$\times$SU(2)$\times$U(1) are antiperiodic and
form a Kaluza-Klein tower starting at the mass level $1/R$. 
For a 4D observer, these boundary conditions break SU(5) down to the Standard Model (SM) gauge group
at a GUT scale of order $1/R$. 
The doublet-triplet splitting problem is solved if the parity $\Omega$ is extended to the Higgs multiplets $H_5$ and $H_{\overline{5}}$,
also assumed to live in the bulk, in such a way that the electroweak doublets are periodic (and then have zero modes),
whereas the colour triplets are antiperiodic (getting masses of order $1/R$).
To reduce $N=2$ SUSY (induced by the original $N=1$ SUSY in five dimensions) down to $N=1$ it is necessary to compactify the fifth dimension on the orbifold 
$S^1/Z_2$ rather than on the circle $S^1$. The orbifold projection eliminates all the zero modes of the extra states belonging to $N=2$ SUSY
and also those of the fifth component of the gauge vector bosons. The zero modes we are left with are the 4D gauge bosons of the
SM, two electroweak doublets and their $N=1$ SUSY partners. 
For the gauge vector bosons and the Higgses we adopt this setup, which is described in detail in refs. \cite{5D}.
For the remaining fields we have much more freedom \cite{5D,5Dfreedom}: they can be located  in the bulk, at the SU(5) preserving brane $y=0$, or at the SU(5) breaking brane $y=\pi R$. 
In our construction based on $S_4$, the three $\bar{5}$ are grouped into the $S_4$ triplet $F$, while
the tenplets $T_1$, $T_2$ and $T_3$ are assigned to the singlet $1$ of $S_4$.
We choose to put the tenplet of the first two families $T_1,T_2$ in the bulk and all remaining $N=1$ supermultiplets on the SU(5) preserving brane at $y=0$.
To obtain the correct zero mode spectrum with intrinsic parities compatible with symmetry and orbifolding, one must introduce two copies of each 
multiplet with opposite parity $\Omega$ in the bulk. 
Therefore $T_{1,2}$ is a short notation for the copies $T_{1,2}$ and $T_{1,2}^\prime$; the zero modes of $T_{1,2}$ are the SU(2) quark doublets $Q_{1,2}$,
while those of $T_{1,2}'$ are $U_{1,2}^c$ and $E_{1,2}^c$.  This setup has a two-fold advantage: an automatic suppression of the Yukawa couplings 
for the fields living in the bulk and the breaking of the  mass relation $m_e = m_d^T$ for the first two fermion families\footnote{It is interesting to observe that, to break the SU(5) relation $m_e = m_d^T$ in the case $T_{1,2}$ are also localized on the brane, 
one could introduce a ${\overline {45}}$ representation for the Higgs fields,
also propagating in the bulk. However, it turns out that this representation not only contains zero modes for the electroweak doublet but 
also for colored Higgses, thus reintroducing a sort of doublet-triplet splitting problem.}. The latter is a consequence 
of using the copies $T_{1,2}$ and $T^\prime_{1,2}$ for the down quarks and charged leptons, respectively,
whereas the former originates from the fact that a bulk field $B$ and its zero mode $B^0$ are related by:
\be
B=\frac{1}{\sqrt{\pi R}} B^0+...
\ee
where dots stand for the higher modes. This expansion produces a (geometrical) suppression factor
\be
s\equiv\dd\frac{1}{\sqrt{\pi R \Lambda}}<1 
\label{vsup}
\ee  
entering the Yukawa couplings depending on the field $B^0$. 
This applies also to the Higgs vevs; since all the matter fields (but $T_{1,2}$) are localized at $y=0$, what matter for the Yukawa couplings are the 
values of the vevs  at $y=0$:
\bea
\langle H_5 (0)\rangle =\frac{v_u^0}{\sqrt{\pi R}}  \qquad \langle  H_{\overline 5} (0)\rangle =\frac{v_d^0}{\sqrt{\pi R}}  
\,.
\eea
A similar setup (to which we refer for further details) has been used in \cite{Altarelli:2008bg} in the context of a SUSY SU(5)$\times A_4$. 
To break the $S_4$ symmetry, we consider a set of  SU(5)-invariant flavon
supermultiplets: three triplets $\varphi_\ell, \varphi_\nu$ ($3_1$), $\chi_\ell$ ($3_2$) and 
one singlet $\xi_\nu$. The alignment of their vacuum expectation values (vevs) along appropriate directions in
flavour space will be the source of BM lepton mixing. For this to work, we employ a cyclic $Z_3$ symmetry 
which allows the fields $\varphi_\nu$ and $\xi_\nu$ to be the only ones responsible for neutrino masses at leading order.
The GUT Higgs fields $H^{}_{5}$ and $H_{\overline{5}}$ are singlets under the family symmetry but charged in the same way under $Z_3$,
so that they are distinguished only by their SU(5) transformation properties.
To  achieve a realistic mass spectrum, beside the geometrical suppression factors, we also exploit the Froggatt-Nielsen mechanism.
The tenplets $T_1$ and $T_2$ are charged under a $U(1)_{FN}$ flavour group, spontaneously broken by the vevs of two fields
$\theta$ and $\theta^\prime$ both carrying $U(1)_{FN}$ charges $-1$ and transforming as a singlet of $S_4$. 
The assignment of the fields under SU(5) and the discrete group $S_4 \times Z_3$ is 
summarized in Tab.\ref{tab:Multiplet1}.
\begin{table}[h!]
\begin{center}
\begin{tabular}{|c||c|c|c|c|c|c|c|c|c|c|c|c|c|c|c|c||}
\hline
{\tt Field} & $F$ & $T_1$ & $T_2$ & $T_3$ & $H_5$ & $H_{\overline 5}$ &   $\varphi_\nu$ & $\xi_\nu$ &  $\varphi_\ell$& $\chi_\ell$ &  $\theta$ &  $\theta^\prime$ & 
$\varphi^0_\nu$& $\xi^0_\nu$ &  $\psi^0_\ell$& $\chi^0_\ell$ 
\\
\hline\hline
SU(5) & $\bar{5}$  & 10 & 10 & 10 & 5 & ${\overline {5}}$ &  1 & 1 & 1 & 1 & 1 & 1 & 1 & 1 & 1 & 1  \\
\hline
$S_4$  &  $3_1$ & 1  & 1 & 1 & 1 & 1 &   $3_1$ & 1 & $3_1$ & $3_2$ &  1 & 1 & $3_1$ & 1 & 2 & $3_2$     \\
\hline
$Z_3$ & $\omega$ & $\omega$ &1  &  $\omega^2$ & $\omega^2$ &$\omega^2$ &  1 & 1 & $\omega$ &$\omega$ &1 &  $\omega$ & 1 & 1 & $\omega$ & $\omega$  \\
\hline
$U(1)_R$ & 1 & 1 & 1 & 1 & 0 & 0&  0 & 0 & 0 & 0& 0 & 0 & 2 & 2 & 2 & 2 \\
\hline
$U(1)_{FN}$ & 0 & 3 & 1 & 0 & 0 &  0 & 0 & 0 & 0& 0 & -1 & -1 & 0 & 0 & 0 & 0 \\
\hline 
  & {\tt br} &  {\tt bu} & {\tt bu} & {\tt br} &  {\tt bu} &  {\tt bu} &   {\tt br} &  {\tt br} &  {\tt br} & 
 {\tt br} &  {\tt br} &  {\tt br} &  {\tt br} &  {\tt br} &  {\tt br} &  {\tt br}\\
\hline 
\end{tabular}
\caption{\label{tab:Multiplet1}\it Matter assignment of the model. The symbol ${\tt br}({\tt bu})$ indicates that the corresponding fields live
on the brane (bulk).}
\end{center}
\end{table}
Before closing this section, it is interesting to outline that, as long as the U(1)$_R$ symmetry remains unbroken, 
dangerous operators that could spoil the solution of the doublet-triplet splitting problem or mediate proton decay are forbidden to all orders. 
In fact, given the  U(1)$_R$ assignments in Tab.\ref{tab:Multiplet1},  the mass term $H_5 H_{\bar 5}$ has  U(1)$_R=0$ and cannot be included in the 
superpotential    of the effective $N=1$ SUSY, which  should have U(1)$_R$ charge $+2$, to compensate the $R$-charge $-2$ coming
from the Grassmann integration measure $d^2\theta$. Also, all renormalizable baryon and lepton number violating operators, such as $F H_5$ and $F F T$, are not 
allowed, and the dimension five operator $FTTT$, leading to proton decay, has $R$-charge $+4$ and therefore is absent. 
The paper is organized as follows: in Sect.\ref{vacall} we discuss the vacuum alignments of the flavon fields at LO and NLO, since they are the necessary ingredients
to build the mass matrices for charged and neutral fermions; in Sect.\ref{downq} we show how to get
a realistic pattern of the mass ratios in the down quark and charged leptons, also computing the left-handed rotations in both sectors, needed to 
build the CKM matrix and correct the neutrino mixing matrix. Sect.\ref{upq} is devoted to the up-type quarks whereas in sect.\ref{neusect} 
we discuss the neutrino sector and some phenomenological implications of the LO and NLO mass matrices. In Sect.\ref{concl} we draw our conclusions.

\section{Vacuum alignment}
\label{vacall}
We solve the vacuum alignment problem using the method first introduced in
\cite{Altarelli:2005yx}. Within this approach, a  continuous
$U(1)_R$ symmetry is introduced, under which matter fields have $R=+1$, while
Higgses and flavon fields have $R=0$. 
Such a symmetry will be
eventually broken down to the R-parity by small SUSY breaking
effects which can be neglected in the first approximation. The required vacuum alignment is obtained 
introducing the so-called driving fields with $R=+2$, which enter
linearly into the superpotential. 
We use here the same flavon content and $S_4$ property transformations as in \cite{Altarelli:2009gn}, that is two triplets  
$\varphi^0_\nu$ and $\chi^0_\ell$, one doublet $\psi^0_\ell$ and one singlet 
$\xi^0_\nu$.
We assume that the family symmetry is broken at an energy scale where SUSY is still an exact symmetry;
this allows to deduce the alignment of the flavon fields from equations arising from setting 
the F-terms of the driving fields to zero.
All the multiplets but the flavon ones have vanishing 
vevs and set to zero for the present discussion. 
We regard the $U(1)_{FN}$ Froggatt-Nielsen flavour symmetry as a local symmetry, assuming that other vector-like multiplets (not specified here)
are introduced to remove the anomaly associated with  $U(1)_{FN}$.
Within these assumptions the relevant part of the scalar potential of the model is given by the sum of 
 the F-terms and of a D-term:
\be
V=V_F+V_D~~~,
\ee
with 
\be
V_F=\sum_i\left\vert\frac{\partial w}{\partial\varphi_i}\right\vert^2\,.
\ee

\subsection{Leading order}
The LO superpotential responsible for the vacuum aligment is equal to the one quoted in \cite{Altarelli:2009gn}: 
\bea
w_d\;=&&M_\varphi (\varphi_\nu^0\varphi_\nu)+g_1\left(\varphi_\nu^0(\varphi_\nu\varphi_\nu)_{3_1}\right)+g_2\left(\varphi_\nu^0\varphi_\nu\right)\xi_\nu+\nn\\
&+&\xi_\nu^0 \left[ M_\xi^2 +M^\prime_\xi  \xi_\nu+g_3 (\varphi_\nu\varphi_\nu) + g_4 \xi_\nu\xi_\nu \right]+\label{wd}\\
&+&f_1\left(\psi_\ell^0(\varphi_\ell\varphi_\ell)_2\right)+f_2\left(\psi_\ell^0(\chi_\ell\chi_\ell)_2\right) +
f_3\left(\psi_\ell^0(\varphi_\ell\chi_\ell)_2\right)+\nn\\
&+&f_4\left(\chi_\ell^0(\varphi_\ell\chi_\ell)_{3_2}\right) \nn\,.
\eea
We parametrize the triplet flavon vevs as 
\bea
\nonumber
\langle \phi\rangle = v_\phi\left( \begin{array}{c}
                        \phi_1 \\
                        \phi_2\\
                        \phi_3  \\
                     \end{array}\right)\,.
\eea
In the SUSY limit, the vacuum configuration is determined by the
vanishing of the derivative of $w_d$ with respect to each component
of the driving fields. The set of such equations for the minimum of the potential can be divided into two decoupled parts: one for
the neutrino sector (involving $\varphi_\nu$ and $\xi_\nu$) and one for the charged lepton sector (driven by $\varphi_\ell$ and $\chi_\ell$). ù
We do not report here such equations (since they are equal to \cite{Altarelli:2009gn}) and only quote 
the final results:
\bea
\nn
\langle\varphi_\ell\rangle &=&v_{\varphi_\ell} \left(
                     \begin{array}{c}
                       0 \\
                       1 \\
                       0 \\
                     \end{array}
                   \right) \qquad\qquad
                   \langle \chi_\ell\rangle=v_{\chi }\left(
                     \begin{array}{c}
                       0 \\
                       0 \\
                       1 \\
                     \end{array}
                   \right) \\ \label{vev:neutrinos} \\
\langle\varphi_\nu\rangle& =&v_{\varphi_\nu} \left(
                     \begin{array}{c}
                       0 \\
                       1 \\
                       -1 \\
                     \end{array}
                   \right) \qquad\quad
\langle\xi_\nu\rangle = v_{\xi}\nn
\eea
where the various $v_\phi$'s obey: 
\bea
\label{conditions}
2 g_3 v_{\varphi_\nu}^2 &=&  \frac{g_4 M^2_\varphi}{g_2^2}-\frac{M_\xi^\prime M_\varphi }{g_2}+M_\xi^2 \nn \\
\sqrt{3} f_1 v_{\varphi_\ell}^2&=&-v_\chi  \left(\sqrt{3} f_2 v_\chi+ f_3 v_{\varphi_\ell} \right) \\
v_\xi &=& -\frac{M_\varphi }{g_2}\nn  \,,
\eea
\noindent
with $v_{\chi}$ undetermined.
The D-term is given by:
\be
V_D=\frac{1}{2}(M_{FI}^2- g_{FN}\vert\theta\vert^2-g_{FN}\vert\theta^\prime\vert^2+...)^2
\ee
where $g_{FN}$ is the gauge coupling constant of $U(1)_{FN}$ and $M_{FI}^2$ denotes the contribution of the Fayet-Iliopoulos term. 
We have omitted the SU(5) contribution to the D-term, whose vev is 
 zero.
There are SUSY minima such that $V_F=V_D=0$. The vanishing of $V_D$
requires
\be
g_{FN}\vert\theta\vert^2+g_{FN}\vert\theta^\prime\vert^2=M_{FI}^2~~~.
\ee
If the parameter $M_{FI}^2$ is positive, the above condition
determines a non-vanishing vev for a combination of $\theta$ and
$\theta^\prime$.  

\subsection{Next to leading order}

The next level of approximation is different from \cite{Altarelli:2009gn}. 
The corrections to the flavon superpotential can be expressed as follows:
\bea
\Lambda \Delta w_{d}=\sum_{i=1}^{15}\alpha_i I_i^{\varphi_\nu^0}+
\sum_{i=1}^5 \beta_i I_i^{\xi_\nu^0}+\sum_{i=1}^7 \gamma_i I_i^{\psi_l^0}+
\sum_{i=1}^{9} \delta_i I_i^{\chi_\ell^0} \,,
\eea
where the following operators contribute to the quartic invariants 
$\left\{I_i^{\varphi_\nu^0},\;I_i^{\xi_\nu^0},\;I_i^{\psi_l^0},\;I_i^{\chi_\ell^0}\right\}$ 
(we do not specify here the different contractions 
among the fields, see appendix A for details):
\bea
I_i^{\varphi_\nu^0}&:& \qquad  \varphi_\nu^3, \; \varphi_\ell^3, \; \chi_\ell^3,\;\varphi_\nu^2\xi_\nu,\;\varphi_\ell^2\chi_\ell,\;
 \chi_\ell^2\varphi_\ell, \; \varphi_\nu\xi_\nu^2 \nn \\
I_i^{\xi_\nu^0}&:& \qquad \varphi_\nu^3, \;\xi_\nu^3,\; \varphi_\ell^3, \; \varphi_\nu^2\xi_\nu,\;
\chi_\ell^2\varphi_\ell \label{vevNLO} \\
I_i^{\psi_l^0}&:& \qquad \varphi_\ell^2\varphi_\nu,\;\varphi_\ell^2\xi_\nu,\;\chi_\ell^2\varphi_\nu,\;
\chi_\ell^2\xi_\nu,\;\varphi_\nu\varphi_\ell\chi_\ell,\;\xi_\nu\varphi_\ell\chi_\ell\nn \\
I_i^{\chi_\ell^0}&:&\qquad\varphi_\ell^2\varphi_\nu,\;\chi_\ell^2\varphi_\nu,\;\varphi_\nu\varphi_\ell\chi_\ell,\;\xi_\nu\varphi_\ell\chi_\ell \nn\,.
\eea
The relevant feature of such corrections is the presence of terms where flavons of the neutrino and charged lepton sectors mix to each other.  
To get the NLO vacua, we parametrize the vev shifts according to:
\bea
\langle\varphi_\ell\rangle &=&\label{vev:charged:best}v_{\varphi_\ell} \left(
                     \begin{array}{c}
                       \delta_{\varphi_1} \\
                       1+ \delta_{\varphi_2}\\
                       \delta_{\varphi_3} \\
                     \end{array}
                   \right) \qquad\qquad
                    \langle\chi_\ell\rangle=v_{\chi }\left(
                     \begin{array}{c}
                       \delta_{\chi_1} \\
                       \delta_{\chi_2} \\
                       1 \\
                     \end{array}
                   \right) \\
\langle\varphi_\nu\rangle& =&v_{\varphi_\nu} \left(
                     \begin{array}{c}
                        \delta_{\nu_1} \\
                       1+\delta_{\nu_2} \\
                       -1+ \delta_{\nu_3}\\
                     \end{array}
                   \right)  \qquad ~~~~\langle\xi_\nu\rangle =v_{\xi} (1 + \delta v_\xi)
\label{vev:neutrinos2}
\eea
where all $\delta_i$ are smaller than 1, and look for solutions by imposing the vanishing of the first 
derivative of $w_d+\Delta w_d$ with respect to the driving fields. After some algebra, we get the following results:
\bea
\delta_{\varphi_1}&\ne&0  \qquad \delta_{\varphi_2}\ne0  \qquad \delta_{\varphi_3} =  0 \\
\delta_{\chi_1}&\ne& 0  \qquad  \delta_{\chi_2} = 0 \\
\delta_{\nu_1} &\ne& 0 \qquad \delta_{\nu_2} = -  \delta_{\nu_3}\,,
\eea
where all the non-vanishing $\delta_i$ are proportional to a combination of $v_i/\Lambda$; this ratio also fixes the relative magnitude
between the leading order components of the flavon vevs and the  $\delta_i$'s.  To make this more transparent, 
 we rescale the  perturbations $\delta_i \to \varepsilon^\prime \,\delta_i$, with now $\delta_i\sim {\cal O}(1)$.
It is important to observe that the corrections to the second and third component of $\varphi_\nu$ are opposite to each other and can be reabsorbed into 
the leading order results. The same happens for the non-vanishing corrections to the second component of $\varphi_\ell$.
According to what discussed, the flavon vev structure that will be used to determine the fermion masses can be summarized as follows:
\bea
\langle\varphi_\ell\rangle &=&\label{vev:charged:best2}v_{\varphi_\ell} \left(
                     \begin{array}{c}
                       \varepsilon^\prime\, \delta_{\varphi_1} \\
                       1 \\
                       0 \\
                     \end{array}
                   \right) \qquad
                   \langle \chi_\ell\rangle=v_{\chi }\left(
                     \begin{array}{c}
                       \varepsilon^\prime\,\delta_{\chi_1} \\
                       0 \\
                       1 \\
                     \end{array}
                   \right) \qquad 
\langle\varphi_\nu \rangle=v_{\varphi_\nu} \left(
                     \begin{array}{c}
                       \varepsilon^\prime \,\delta_{\nu_1} \\
                       1 \\
                       -1 \\
                     \end{array}
                   \right)  \,;
\label{vev:neutrinos3}
\eea
The magnitude of the flavon vevs as well as of the $\varepsilon^\prime$ perturbations by will be discussed in the next section.

\section{Down quarks and charged lepton mass matrices}
\label{downq}
The superpotential built with operators with two-flavon insertions (beside the flavons carrying Froggatt-Nielsen charges) 
allows to determine the relevant features of the down quarks and charged lepton mass matrices. 
According to the discussion in the Introduction, we use the same notation for 
the couplings involving the tentplet $T_{1,2}$, although it  should be understood that they are different for down quarks and 
charged leptons. The superpotential in this sector reads as follows:
\bea
w_\ell &=& 
 F T_3 H_{\overline 5}  \left[\frac{\alpha_b}{\Lambda^{3/2}}\varphi_\ell +
\frac{\alpha_1}{\Lambda^{5/2}}  (\varphi_\nu \varphi_\ell)_{3_1} +
\frac{\alpha_2}{\Lambda^{5/2}}   (\varphi_\nu \chi_\ell ) _{3_1}+  \frac{\alpha_3}{\Lambda^{5/2}}  \varphi_\ell \xi_\nu +
\right] +
 \nn \\ \nn \\
&&  F T_2 H_{\overline 5} \,\theta \left[\frac{\beta_1}{\Lambda^{3}}   \varphi_\nu + 
\frac{\beta_2}{\Lambda^{4}}   (\varphi_\nu^2)_{3_1} 
+ \frac{\beta_3}{\Lambda^{4}}   \varphi_\nu \xi_\nu\right] + 
  \nn \\ \nn \\
&&  F T_2 H_{\overline 5} \,\theta^\prime \left[
\frac{\beta_4}{\Lambda^{4}}   (\varphi_\ell^2)_{3_1} 
+ \frac{\beta_5}{\Lambda^{4}} (\chi_\ell^2)_{3_1}   + 
\frac{\beta_6}{\Lambda^{4}}   (\chi_\ell \varphi_\ell ) _{3_1} \right]+
  \nn \\ \nn \\
 &&  F T_1 H_{\overline 5} \, \theta^3 \left[
\frac{\gamma_1}{\Lambda^{4}} (\varphi_\ell^2)_{3_1}+
\frac{\gamma_2}{\Lambda^{4}} (\chi_\ell^2)_{3_1} +\frac{\gamma_3}{\Lambda^{4}} (\varphi_\ell\chi_\ell)_{3_1}\right]+  \label{down}
\\ \nn \\
&& F T_1 H_{\overline {5}}\, \theta^2 \theta^\prime \left[\frac{\gamma_4}{\Lambda^{3}}   \varphi_\ell
+ \frac{\gamma_5}{\Lambda^{4}}   (\varphi_\nu\varphi_\ell)_{3_1}+
 \frac{\gamma_6}{\Lambda^{4}}   (\varphi_\nu\chi_\ell)_{3_1}+\frac{\gamma_7}{\Lambda^{4}} \xi_\nu  \varphi_\ell\right] \nn+  \\ \nn \\
&& F T_1 H_{\overline {5}}\, \theta \theta^{\prime 2} \left[\frac{\gamma_8}{\Lambda^{3}}   \varphi_\nu
+ \frac{\gamma_9}{\Lambda^{4}}   (\varphi_\nu^2)_{3_1}+
\frac{\gamma_{10}}{\Lambda^{4}} \xi_\nu  \varphi_\nu\right]+
\nn \\ \nn \\
&& F T_1 H_{\overline {5}}\,  \theta^{\prime 3} \left[ \frac{\gamma_{11}}{\Lambda^{4}} (\varphi_\ell^2)_{3_1}+
\frac{\gamma_{12}}{\Lambda^{4}} (\chi_\ell^2)_{3_1} +\frac{\gamma_{13}}{\Lambda^{4}} (\varphi_\ell\chi_\ell)_{3_1}\right]\,,\nn
\eea
where we took into account that the fields living in the bulk have mass dimension 3/2.
From the superpotential in eq.(\ref{wd}) we expect a common order of magnitude for the vev's (scaled by the cutoff $\Lambda$):
\bea
v_{\varphi_\ell} \sim v_{\chi} \sim \varepsilon \qquad v_{\varphi_\nu} \sim v_{\xi} \sim \epsilon\;.
\eea
although, due to the different minimization conditions that determine $(v_{\varphi_\ell},v_{\chi} )$ and $(v_{\varphi_\nu},v_{\xi})$, we may tolerate a moderate hierarchy
between $\varepsilon$ and $\epsilon$. Similarly the order of magnitude of $\langle \theta\rangle$ and $\langle \theta^\prime \rangle$ is in principle unrelated to those of $\varepsilon$ and $\epsilon$.
In the following, we assume that $\varepsilon = \epsilon$ and use
the short-hand notations for the ratio among flavon vevs and $\Lambda$:
\bea
\label{definition}
\frac{v_{\varphi_\ell}}{\Lambda} =\frac{v_{\chi}}{\Lambda} = \frac{v_{\varphi_\nu}}{\Lambda} = \frac{v_{\xi}}{\Lambda} = \varepsilon 
\qquad
\frac{\langle \theta \rangle}{\Lambda} = t \qquad \frac{\langle \theta^\prime \rangle}{\Lambda} = t^\prime \,.
\eea
Disregarding for the moment all ${\cal O}(1)$ coefficients and taking the vev $\langle H_{\overline 5}\rangle=
v_d$, 
the mass matrix (in the $\bar \psi_L m_d \psi_R$ convention) obtained from this Lagrangian is:
\bea
\label{massLO}
m_d = v_d s\, \varepsilon 
\left(
\begin{array}{ccc}
s \,\varepsilon\, (t^3 + t^2 t^\prime+t^{\prime 3})  & s\,t t^{\prime 2}& s\,t t^\prime (t+t^\prime) \\
s \varepsilon\,t^\prime                   & s t&  s t   \\
   \varepsilon                    & 0 & 1 \\
                     \end{array}
                   \right)\,,
\eea
where $s$ is the suppression volume factor. 
If $\varepsilon,t$ and $t^\prime$ are relatively close in magnitude,
we can easily recognize that the down quark and charged lepton mass hierarchies are given by:
\bea
m_b : m_s : m_d \sim m_\tau : m_\mu : m_e  \sim 1: s t  : s \,\varepsilon\, (t^3 + t^2 t^\prime+t^{\prime 3})
\eea
whereas the mixing angles in the down sector can be estimated to be:
\bea
\theta_{12}^d \sim  t^{\prime 2}\qquad \theta_{13}^d \sim s\,t t^\prime (t+t^\prime)  \qquad \theta_{23}^d \sim s t\,.
\eea
Notice that, by transposition, we also get an estimate of the charged lepton mixings, given by:
\bea
\theta_{12}^\ell \sim \varepsilon\,\left(\frac{t^\prime}{t}\right) \qquad \theta_{13}^\ell \sim \varepsilon \qquad \theta_{23}^\ell \sim 0.
\eea
Then, a realistic pattern of fermion masses and mixings can be achieved requiring that
\bea
\label{realistic}
s \sim \varepsilon \sim  t\sim t^\prime \sim \lambda\,,
\eea
where $ \lambda \equiv \lambda_C$. Given that $\varepsilon^\prime\,\delta_i \sim v_i /\Lambda$, we can also fix $\varepsilon^\prime\sim \lambda$.
Before discussing in details the mass matrices obtained in such a situation, it is useful to consider the relevant
corrections coming from three-flavon insertion operators and from the vev shifts quoted in 
eq.(\ref{vev:neutrinos3}). The relevant feature of such corrections is the filling of the vanishing element of the mass matrix in eq.(\ref{massLO}), which turns
out to be $s\,\varepsilon^3$, that is of ${\cal O}(\lambda^4)$. This allows to shift $\theta_{23}^\ell$ away from zero by a quantity of ${\cal O}(\lambda^2)$.
The relevant operators are of the form  
$F T_3 H_{\bar 5}\,(\varphi_\ell \varphi_\nu^2+\chi_\ell\varphi_\nu^2)$ computed with leading order vevs and 
$F T_3 H_{\bar 5}\,(\varphi_\nu\varphi_\ell+\varphi_\nu\chi_\ell)$ with flavon vevs at the next to leading order (see appendix B for details). 
Putting together all these elements we 
are now in the position to write down the most general mass matrix allowed in our model:
\bea
\label{massNLO}
m_d \sim v_d  \, \lambda^2
\left(
\begin{array}{ccc}
\lambda^5  & \lambda^4 & \lambda^4  \\
\lambda^3  &  \lambda^2 & \lambda^2  \\
\lambda   &  \lambda^2   & 1 \\
                     \end{array}
                   \right)\,.
\eea
The unitary left-handed rotation $U_d$ is obtained diagonalizing $m_d\,m_d^\dagger$ whereas the right-handed one $U_\ell$
is the charged lepton rotation. In terms of the Cabibbo angle they are given by:
\bea
U_d \sim  
\left(
\begin{array}{ccc}
1  & \lambda^2 & \lambda^4  \\
\lambda^2  &  1 & \lambda^2  \\
\lambda^4   &  \lambda^2   & 1 \\
                     \end{array}
                   \right)
\qquad
U_\ell \sim  
\left(
\begin{array}{ccc}
1  & \lambda & \lambda \\
\lambda  &  1 & \lambda^2  \\
\lambda   &  \lambda^2   & 1 \\
                     \end{array}
                   \right)\,.
\qquad
\eea 
We see that $U_d$ perfectly reproduces the correct order of magnitude of $V_{ub}$ and $V_{cb}$, whereas $V_{us}$ turns out to be a bit smaller
than expected, as in many other models based on non-abelian discrete symmetries. 
On the other hand,  in the construction of the neutrino mixing matrix $U_{PMNS}$ \cite{Pontecorvo:1957cp}, $U_\ell$ will induce  corrections 
of  ${\cal O}(\lambda)$ to $\theta_{12}$ and $\theta_{13}$ and  ${\cal O}(\lambda^2)$ to $\theta_{23}$. These corrections
turns out to be relevant to shift the solar and reactor angles from their BM values to the experimental ones.
 
For the sake of completeness, we recompute the previous quantities including all ${\cal O}(1)$ coefficients. The mass matrix of the down quarks is now given by:
\bea
\label{massNLO2}
m_d \sim \alpha_b \,v_d\, \lambda^2
\left(
\begin{array}{ccc}
x_1\lambda^5  & x_2\lambda^4 & x_3\lambda^4  \\
x_4\lambda^3  &  x_5\lambda^2 & -x_5\lambda^2  \\
x_6 \lambda   &  x_7\lambda^2   & 1 \\
                     \end{array}
                   \right)\,,    
\eea
where $\alpha_b$ is the bottom Yukawa coupling from the first operator in eq.(\ref{down}) and $x_i$ are linear combinations of the other Yukawa couplings 
of eq.(\ref{down}), rescaled by $\alpha_b$:
\bea 
x_1 &=&  -\gamma _1 + \gamma_2 - \gamma_3 - \gamma_5 - \gamma_6 - \gamma_{11} +\gamma_{12} - \gamma_{13} + \gamma_8 \delta_{\nu_1} + 
\gamma_4 \delta_{\varphi_1}\nn \\
x_2 &=& -\gamma _8
 \nn \\
x_3&=& \gamma _4+\gamma _8\nn \\
x_4 &=&  -\beta_4 + \beta_5 - \beta_6 + \beta_1 \delta_{\nu_1}  \\
x_5 &=&  -\beta_1 \nn \\
x_6 &=&  -\alpha_1 - \alpha_2 + \alpha_b \delta_{\varphi_1} \nn \\
2 x_7 &=&  3 \alpha_5 + \sqrt{3} \alpha_7 - 
 2 \alpha_1 (\delta_{\nu_1} + \delta_{\varphi_1}) + 
 2 \alpha_2 (\delta_{\nu_1} + \delta_{\chi_1}) \nn \,.
\eea
Notice that the (22) and (23) elements are opposite to each other; this equality is broken only at ${\cal O}(\lambda^6)$, for example, 
by operators of the form 
$F\,T_2\,\theta\,(\varphi^2_\ell)_2\varphi_\ell$.
From eq.(\ref{massNLO2}) it is easy to obtain the expression for the charged lepton mass matrix by transposition and changing accordingly 
$x_{1}-x_{5}$ with $x_{1}^\prime-x_{5}^\prime$, which differ from the previous ones by ${\cal O}(1)$ coefficients. This is 
the effect of  introducing the copies $T^\prime_{1,2}$ of the first two tenplet fields, whose zero modes are different from those of $T_{1,2}$ and couple 
with the charged leptons only.
Summarizing, the mass matrix of the charged leptons is:
\bea
\label{massNLO22}
 m_e \sim \alpha_b  \,v_d\, \lambda^2
 \left(
 \begin{array}{ccc}
 x_1^\prime  \lambda^5  & x_4^\prime \lambda^3 &  x_6\lambda  \\
 x_2^\prime \lambda^4  &  x_5^\prime \lambda^2 & x_7\lambda^2  \\
 x_3^\prime \lambda^4   &  -x_5^\prime \lambda^2   & 1 \\
                      \end{array}
                    \right)\,.  
\eea
Working for simplicity in the limit of real coefficients, the quark and charged lepton masses in unit of $\alpha_b  \,v_d \, \lambda^2$ are explicitly given by:
\bea
\label{massesdown}
m_b &=&  m_\tau   \nn \\
m_s &=& |x_5|\,\lambda^2 \qquad m_\mu = |x_5^\prime|\,\lambda^2\\
m_d&=&  \left|x_1-(x_2+x_3)x_6 -\frac{x_2  x_4}{x_5}\right|\,\lambda^5 \qquad 
m_e=  \left|x_1^\prime-(x_2^\prime+x_3^\prime)x_6 -\frac{x_2^\prime  x_4^\prime}{x_5^\prime}\right|\,\lambda^5 \nn\,.
\eea
We see that, at the GUT scale, the $b-\tau$ unification is recovered (up to very small corrections of ${\cal O}(\lambda^4)$ not listed here); also, the 
other two mass ratios $m_d/m_e$ and $m_s/m_\mu$ are both  of the same order of magnitude but not strictly equal because of different ${\cal O}(1)$ coefficients.

Keeping only the leading order for each matrix elements, the left-handed rotation $U_l$ for charged leptons is given by:
\bea
\label{ul}
U_l = 
\left(
\begin{array}{ccc}
1   &  \left(\frac{x_4^\prime}{x_5^{\prime}}+x_6\right) \lambda&  x_6\lambda  \\
-\left(\frac{x_4^\prime}{x_5^{\prime}}+x_6\right) \lambda &  1 & x_7\lambda^2  \\
-x_6\lambda   &  - \left(x_6^2+x_7+\frac{x_4^\prime x_6}{x_5^\prime}\right) \lambda^2  & 1 \\
                     \end{array}
                   \right)
\eea
whereas that for down quarks $U_d$ is:
\bea
\label{ud}
U_d =  
\left(
\begin{array}{ccc}
1   &  \left(\frac{x_2}{x_5}\right) \lambda^2&  x_3\lambda^4  \\
-\left(\frac{x_2}{x_5}\right) \lambda^2  &  1 & -x_5\lambda^2  \\
-(x_2+x_3) \lambda^4    &  x_5 \lambda^2  & 1 \\
                     \end{array}
                   \right)\,.
\eea

\section{Up quarks  mass matrix}
\label{upq}
The up quark mass matrix is completely determined by operators with no more than two-flavon insertions. In the following
we  list the relevant non-vanishing terms in the superpotential disregarding those flavon contractions which will give 
contributions only at the NLO, not relevant for our discussion:
\bea
w_{up} &=& \frac{\alpha_t}{\Lambda^{1/2}} T_3 T_3 H_5 + 
\frac{\delta}{\Lambda^{4}} T_2 T_3 H_5 \theta^\prime (\varphi_\nu\varphi_\ell)
+\frac{\sigma}{\Lambda^{4}} T_1 T_3 H_5 
\theta^2\theta^\prime +
\nn \\ \nn \\ 
&&\frac{1}{\Lambda^{11/2}} T_1 T_2 H_5 \left(\tau_1 \theta^4 + \tau_2 \theta\theta^{\prime 3}\right)+ \\ \nn \\ &&
\frac{\rho}{\Lambda^{7/2}} T_2 T_2 H_5 \theta\theta^{\prime} +
\frac{1}{\Lambda^{15/2}} T_1 T_1 H_5 (\eta_1 \theta^4 \theta^{\prime 2} + \eta_2 \theta\theta^{\prime 5}) \nn\,.
\eea
Taking into account the conditions in eq.(\ref{realistic}), the corresponding mass matrix reads:
\bea
\label{massup}
m_{up} =  v_u^0 \, \alpha_t \, \lambda
\left(
\begin{array}{ccc}
(\eta_1+\eta_2) \lambda^8  & (\tau_1+\tau_2) \lambda^6 & \sigma \lambda^4  \\
(\tau_1+\tau_2) \lambda^6 &   \rho\lambda^4 & -\delta  \lambda^4  \\
\sigma \lambda^4   & -\delta  \lambda^4   & 1 \\
                     \end{array}
                   \right)\,    
\eea
where, once again, we have rescaled the Lagrangian coefficients by $\alpha_t$ and used the same symbols. It is easy now to read masses and mixing matrix;
for the first we have:
\bea
m_{t}:m_c:m_{u} \sim 1:  \lambda^4 :  \lambda^8 
\eea
which perfectly matches with the GUT expectations. To avoid large dimensionless coefficients,  
we assume here that $v_{u,d}\approx \lambda v_{u,d}^0$. We are allowed to do that because of the freedom related to
the boundary values $v_{u,d}^0$; in fact, the electroweak scale is determined by the relations:
\bea
v_u^2+v_d^2\approx (174~{\rm GeV})^2\qquad v_u^2\equiv\int_0^{\pi R} dy \left\vert\langle H_5(y) \rangle\right\vert^2\qquad
v_d^2\equiv\int_0^{\pi R} dy \left\vert\langle H_{\bar 5}(y) \rangle\right\vert^2
\eea 
and, unless  $\langle H_{5,\bar{5}}(y) \rangle$ are constant in the fifth coordinate, one would expect $v_{u,d}^0\ne v_{u,d}$. 
With this assumption,  the Yukawa coupling of the top quark
is of order one and, thanks to eqs.(\ref{massesdown}-\ref{massup}), also all the other couplings are of the same order.
Up to ${\cal O}(\lambda^4)$ the matrix diagonalizing  $m_{up}m^\dagger_{up}$ is
\bea 
\label{uleft}
U_{u} = 
\left(
\begin{array}{ccc}
1  & \left(\frac{\tau_1+\tau_2}{\rho}\right) \lambda^2 &   \sigma \lambda^4   \\
-\left(\frac{\tau_1+\tau_2}{\rho}\right) \lambda^2 &   1 & -\delta\lambda^4  \\
-\sigma \lambda^4   & \delta\lambda^4  & 1 \\
                     \end{array}
                   \right)\,    \,.
\eea
We see that the up sector  contributes to both  $(V_{us}-V_{cd})$ and $(V_{ub}-V_{td})$ of the CKM, which in turn results to be:
\bea
\label{ckm}
V_{CKM}=U_d^\dagger U_{u} =   
\left(
\begin{array}{ccc}
1  & \left(\frac{\tau_1+\tau_2}{\rho}-\frac{x_2}{x_5}\right) \lambda^2 &   \left(\sigma-x_2-x_3\right) \lambda^4   \\
-\left(\frac{\tau_1+\tau_2}{\rho}-\frac{x_2}{x_5}\right) \lambda^2 &   1 & x_5\lambda^2  \\
\left[x_3-\sigma+\frac{x_5}{\rho} (\tau_1+\tau_2)\right] \lambda^4   & -x_5\lambda^2  & 1 \\
                     \end{array}
                   \right)\,    \,.
\eea

It is interesting to observe that the different coefficients of the (13) and (31) elements can explain the experimental 
difference among these matrix elements\footnote{In addition, to reproduce the correct Cabibbo angle with ${\cal O}(1)$ coefficients
one has to ask, for example, for a negative $x_2$ or $x_5$. In this case, it is enough to take the absolute value of these coefficients equal to the unity 
to gain an enhancement of 
a factor of 3.}.
As a final comment of this section, we observe that:
\bea
\frac{m_b}{m_t} \sim \left(\frac{v_d^0}{v_u^0}\right)\,\left(\frac{\alpha_b}{\alpha_t}\right)\,\lambda
\eea
which can easily reproduce the experimental value $ m_b/m_t \sim\lambda^2$ for $v_u^0/v_d^0\sim 1/\lambda$ (and ${\cal O}(1)$ Yukawa couplings).
Considering the scaling $v_{u,d}\approx \lambda v_{u,d}^0$, the condition $v_u^0/v_d^0\sim 1/\lambda$ implies $\tan \beta \sim 5$.

\section{The neutrino sector}
\label{neusect}
In this section we show how to get a description of the neutrino masses and mixings based on the $S_4$ symmetry. As already outlined in the Introduction, 
the mixing matrix at leading order will have the BM structure. 
Our main issue here is to show that dimension 5 Weinberg operators are enough to get the BM mixing matrix at LO 
and a mass spectrum compatible with the data.
Since a see-saw version (with subdominant D=5 contributions) has been carefully studied in \cite{Altarelli:2009gn}, in the second part of this section we limit 
ourselves to show that such a mechanism can also be successfully employed in the GUT version of the model, with identical LO results. 

\subsection{The neutrino sector from effective operators only}
We start writing the leading and next to leading order contributions to neutrino masses from the Weinberg operator:
\bea
\label{nlow2}
w_w= \frac{y_w}{\Lambda^{2}} (F F)_1 H_5 H_5+\frac{y_{w_1}}{\Lambda^{3}} (F F)_{3_1} H_5 H_5 \varphi_\nu + 
\frac{y_{w_2}}{\Lambda^{3}} (F F)_{1} H_5 H_5 \xi_\nu\,,
\eea
from which the mass matrix is (we use the symbol $\lambda$ according to eqs.(\ref{definition})-(\ref{realistic})):
\bea
m_w=\frac{s^2   (v_u^0)^2}{\Lambda}\left(
          \begin{array}{ccc}
            y_w + y_{w_2}\lambda  &  -y_{w_1}\lambda & -y_{w_1}\lambda\\
            -y_{w_1}\lambda & 0 & y_w + y_{w_2}\lambda \\
            -y_{w_1}\lambda & y_w + y_{w_2}\lambda & 0 \\
          \end{array}
        \right)\,.
\eea
This matrix is exactly diagonalized by BM and the eigenvalues are (replacing $s v_u^0$ with $v_u$):
\bea
m_1 &=& \left(\frac{v_u^2}{\Lambda}\right)  \left[y_w+ (y_{w_2} -\sqrt{2} y_{w_1}) \lambda \right] \nn \\ \label{numasses3}
m_2 &=& \left(\frac{v_u^2}{\Lambda}\right)   \left[y_w+ (y_{w_2} +\sqrt{2} y_{w_1}) \lambda \right]  \\ 
m_3 &=& -\left(\frac{v_u^2}{\Lambda}\right)  \left(y_w+ y_{w_2} \lambda \right) \nn.
\eea
Notice that they satisfy the sum-rule: 
\bea
m_1+m_2+2 m_3=0\,,
\eea 
which allows to reduce the number of independent parameters in the neutrino mass matrix.
We can reparametrize the masses  in terms of two different complex parameters:
\bea
y_w + y_{w_2}  \lambda &=& A  = a\, e^{i \phi_a}\nn \\
\sqrt{2} y_{w_1}  \lambda &=& B= b\, e^{i \phi_b} \qquad a,b >0
\eea
from which:
\bea
|m_1|^2 &=&\left(\frac{v_u^2}{\Lambda}\right)^2  \left(a^2+b^2-2 \,a \,b \,\cos \Delta\right)  \nn \\
|m_2|^2 &=&\left(\frac{v_u^2}{\Lambda}\right)^2   \left(a^2+b^2+2 \,a\, b \,\cos \Delta\right) \label{masses}   \\
|m_3|^2 &=&\left(\frac{v_u^2}{\Lambda}\right)^2   a^2  \,,\nn
\eea
where $\Delta=\phi_a-\phi_b$. Since the solar mass difference is given by:
\bea
\Delta m^2_{sol} &=& |m_2|^2-|m_1|^2=4\,\left(\frac{v_u^2}{\Lambda}\right)^2\, a\,b\, \cos \Delta\,,
\eea
the condition $\cos \Delta >0$ must be fulfilled. This implies that the spectrum is of inverted type because the 
condition $|m_3|>|m_2|$ cannot be satisfied.
We then define the quantity
 \bea
\label{atm}
\Delta m^2_{atm} &=& |m_1|^2-|m_3|^2=  \left(\frac{v_u^2}{\Lambda}\right)^2 b\, (b - 2 a \cos \Delta)
\eea
which is positive for 
\bea
\label{cos}
\cos \Delta < \frac{b}{2a}\,.
\eea
The value of $r$
\bea
r = \frac{\Delta m^2_{sol}}{\Delta m^2_{atm}} = \frac{4 a \cos \Delta}{b-2 a \cos \Delta}
\eea
can be made small only if $\cos \Delta\sim 0$ because the other possible conditions $b>> 2 a \cos \Delta$ reduces to small $\cos \Delta$ and 
$b << 2 a \cos \Delta$ is in conflict with eq.(\ref{cos}). We do not consider here the possibility of $a << 1$ because this parameter
is dominated by $y_w$ and we prefer to work with ${\cal O}(1)$ coefficients in the superpotential.

The experimental value $r\sim 1/30$ is reproduced if 
\bea
\cos \Delta = \frac{b}{2 a (2+r)}\,r \sim 10^{-2}
\eea
for $a\sim b \sim {\cal O}(1)$. Notice that, using this relation into eq.(\ref{atm}), we can estimate the order of magnitude of the scale $\Lambda$, which turns 
out to be:
\bea
\Lambda \sim \frac{ b\, v_u^2}{\sqrt{\Delta m^2_{atm} }}\sim 10^{14} \,{\text GeV}\,,
\eea
for $v_u = 100$ GeV and $\Delta m^2_{atm} =2.4 \times 10^{-3}$ eV$^2$.

At the next to leading order, we have to consider two-flavon insertion operators and the shift of the flavon $\varphi_\nu$ of eq.(\ref{vev:neutrinos2}), that is 
$\delta_{\nu_1}$.
The contributions of the new operators to the superpotential is:
\bea
\delta w_w &=&\frac{y_{w_3}}{\Lambda^4} (F F)_{1} (\varphi_\nu\varphi_\nu)_1 H_5 H_5 +\frac{y_{w_4}}{\Lambda^4} (F F)_{2} (\varphi_\nu\varphi_\nu)_2 H_5 H_5 +
\frac{y_{w_5}}{\Lambda^4} (F F)_{3_1} (\varphi_\nu\varphi_\nu)_{3_1} H_5 H_5 +\nn \\ \\ 
&&\frac{y_{w_6}}{\Lambda^4} (F F)_{1} \xi_\nu^2 H_5 H_5 + \frac{y_{w_7}}{\Lambda^4} (F F \varphi_\nu)_{1} \xi_\nu H_5 H_5 \,, \nn
\eea
with the corresponding mass matrix given by:
\bea
\label{massnnlo}
m_w=\frac{v_u^2}{\Lambda}\left(
          \begin{array}{ccc}
            y_w + y_{w_2}\lambda +y_{w}^\prime \lambda^2  &  -y_{w_1}\lambda -y_{w_7}\lambda^2 & -y_{w_1}\lambda-y_{w_7}\lambda^2 \\
            -y_{w_1}\lambda-y_{w_7}\lambda^2 & \frac{1}{2} (3  y_{w_4} -2  y_{w_1}  \delta_{\nu_1})\lambda^2 & y_w + y_{w_2}\lambda +y_{w}^{''} \lambda^2\\
            -y_{w_1}\lambda-y_{w_7}\lambda^2 & y_w + y_{w_2}\lambda+y_{w}^{''} \lambda^2 & \frac{1}{2} (3  y_{w_4} +2  y_{w_1}  \delta_{\nu_1})\lambda^2\\
          \end{array}
        \right)\,,
\eea
where $y_{w}^\prime = -2 y_{w_3} + y_{w_4} + y_{w_6} $ and $y_{w}^{''}   = 1/2 (-4 y_{w_3} - y_{w_4} + 2 y_{w_6})$. It is easy to show that, in the limit
$\delta_{\nu_1} \to 0$, the mass matrix is still diagonalized by BM. 
For $\delta_{\nu_1} \ne 0$, we observe that  $\theta_{13}^\nu$ and $\theta_{12}^\nu$ are not corrected 
whereas  $\theta_{23}^\nu$ is shifted by a quantity of ${\cal O}(\lambda^2)$:
\bea
\theta_{23}^\nu &=&\frac{\pi}{4} + \left(\frac{y_{w_1}}{2 y_w}\right)\delta_{\nu_1}\lambda^2  \,.
\eea
To get the neutrino mixing matrix, we also have to take into account the corrections coming from the charged leptons, eq.(\ref{ul});
to be consistent up to ${\cal O}(\lambda^2)$, we have to include all ${\cal O}(\lambda^2)$ terms in eq.(\ref{ul}), which mainly affect the diagonal elements 
(the elements of ${\cal O}(\lambda)$ are not modified at this order). Since such corrections are important for $\theta_{23}$, what matter for us are the 
$(U_\ell)_{22}$ and $(U_\ell)_{33}$ elements, which turn out to be $(U_\ell)_{22}= 1-(x_4^\prime + x_5^\prime x_6)^2/(2 x_5^{\prime 2}) \,\lambda^2$ 
and  $(U_\ell)_{33}= 1 - (x_6^2/2) \,\lambda^2$. With this in mind, the neutrino mixing angles can be written as:
\bea
\label{final2}
\theta_{13} &=&  \frac{1}{\sqrt{2} }\,\left(\frac{x_4^\prime}{x_5^\prime}\right)\,\lambda\nn \\ \nn \\
\theta_{23} &=&\frac{\pi}{4} + \left[-\frac{x_4^{\prime 2}}{4\, x_5^{\prime 2}}+\frac{x_6^2}{2}+x_7+\delta_{\nu_1} \left(\frac{y_{w_1}}{2 y_w}\right)\right]\lambda^2 \\ \nn \\
\theta_{12} &=&\frac{\pi}{4} -  \frac{1}{\sqrt{2}}\,\left(\frac{x_4^\prime +2\, x_5^\prime x_6 }{ \,x_5^\prime}\right)\,\lambda \nn.
\eea
We explicitly realized the pattern in eq.(\ref{sinNLO}), with all corrections computed analytically. 
These relations show a certain amount of correlation; for example, the solar angle can be written as:
\bea
\theta_{12} &=&\frac{\pi}{4} -\theta_{13} - \sqrt{2}\, x_6 \,\lambda
\eea
so that we expect a large deviation of $\theta_{12}$ from maximal mixing when $\theta_{13}$ is significantly different from zeros,  unless some
cancellations with the  $x_6$ parameter is at work. This is confirmed by a numerical analysis by treating 
the parameters in  eqs.(\ref{ul}) and (\ref{massnnlo}) as random complex numbers. Since it is crucial that the corrections 
from the charged lepton sectors have coefficients of ${\cal O}(1)$, we restrict the absolute values of the $x_i$  between 1/2 and 2; all 
the other parameters are extracted in the range of  absolute values  between 0 and 3. 
Notice that the results in eqs.(\ref{ul}) and (\ref{massnnlo}) are valid at the GUT scale; however, since the model predicts 
inverted hierarchy, the effect of the running to the SUSY scale can be safely neglected.
\begin{figure}[h!]
\hspace{-1.cm}
\includegraphics[scale=.48]{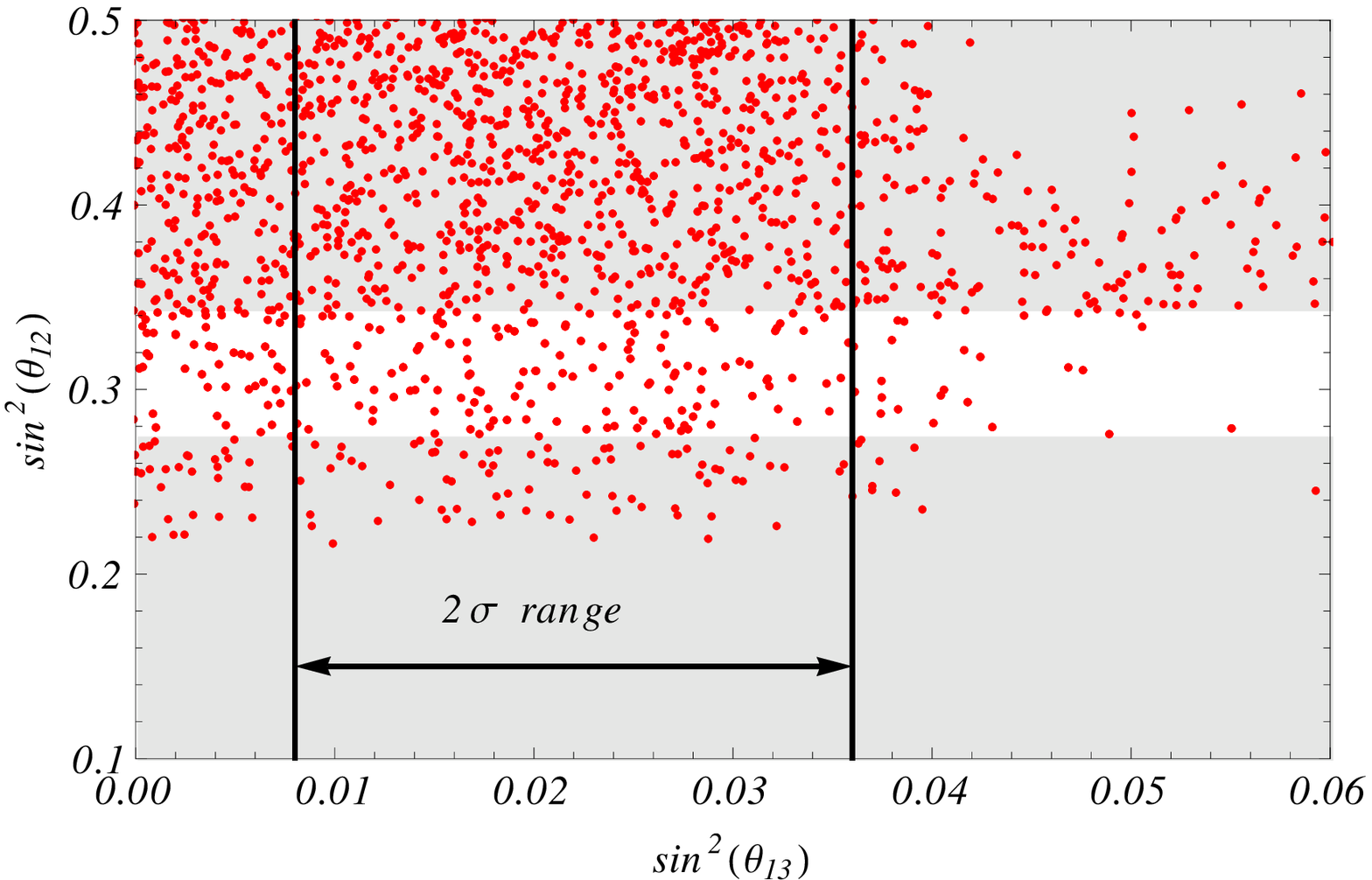}
\qquad
\includegraphics[scale=.48]{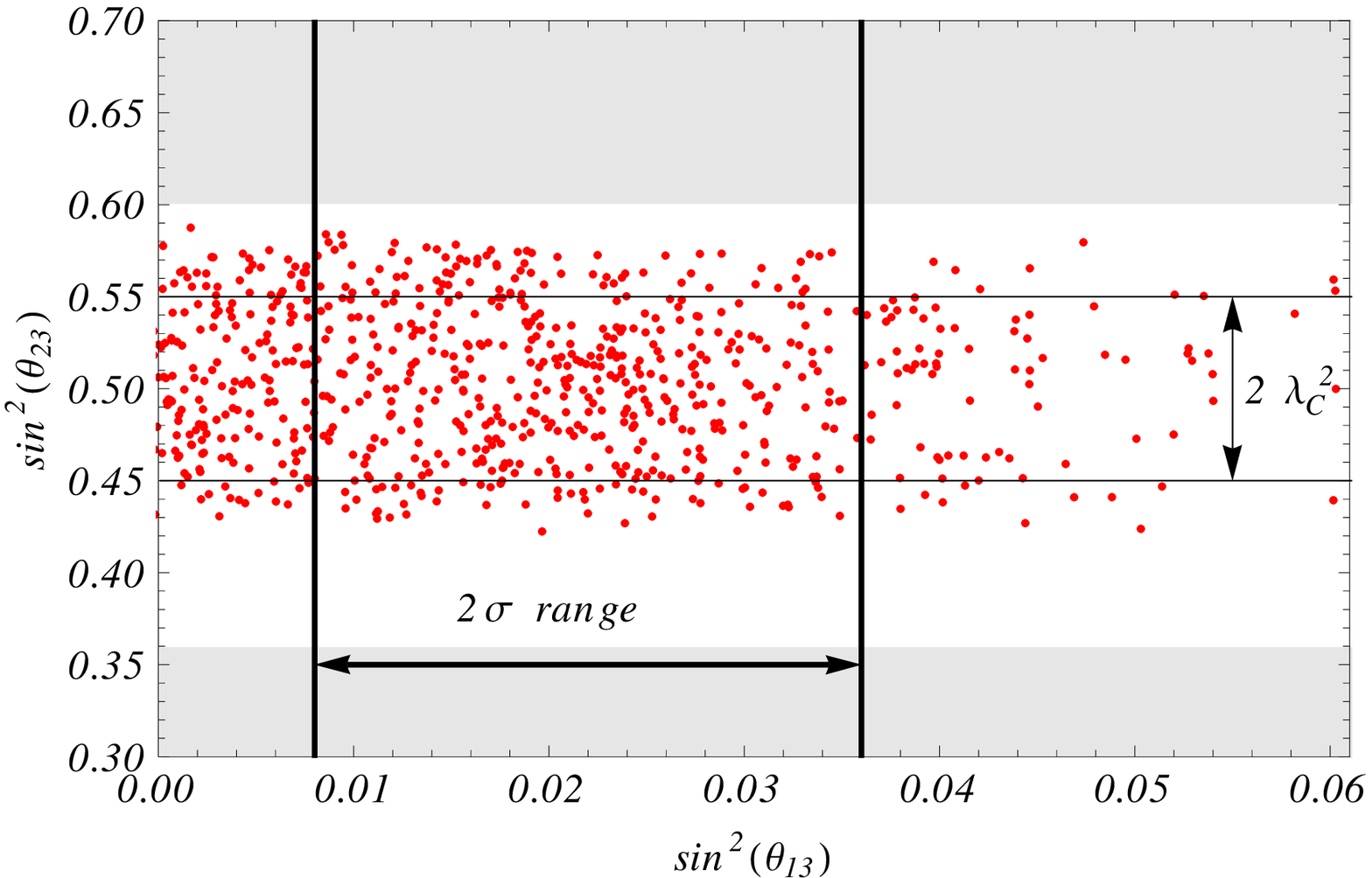}
\caption{\label{fig:corr}\it $\sin^2 \theta_{12}$ as a function of  $\sin^2 \theta_{13}$ (left panel) and
$\sin^2 \theta_{23}$ vs $\sin^2 \theta_{13}$ (right panel).  The gray bands represents the regions excluded by the experimental data at 
2$\sigma$ on $\sin^2 \theta_{12}$ and $\sin^2 \theta_{23}$ \cite{new} whereas
the vertical lines  enclose the $2 \sigma$ range on $\sin^2 \theta_{13}$ as obtained in \cite{new}.  See text for further details.}
\end{figure}
The results of such an analysis are shown in Fig.(\ref{fig:corr}); in the left panel, we plot   $\sin^2 \theta_{12}$ as a function of
$\sin^2 \theta_{13}$. The gray bands represent the regions excluded by the experimental data at 2$\sigma$ on $\sin^2 \theta_{12}$ \cite{new} whereas
the vertical lines enclose the $2 \sigma$ range on $\sin^2 \theta_{13}$ as obtained in \cite{new} (old reactor fluxes). 
The points  correspond  to choices of the parameters reproducing
$\Delta m^2_{atm}$, $\Delta m^2_{sol}$ and the mixing angles within a 3$\sigma$ interval, for the inverted ordering of the neutrino mass spectrum.
We see that a relevant fraction of points fall in the box allowed by the current limits on neutrino parameters. 
In the right plot, we show $\sin^2 \theta_{23}$ (at 2$\sigma$ from \cite{new}) as a function of 
$\sin^2 \theta_{13}$. To guide the eye, we also display an interval of $\pm \lambda_C^2$ around $\sin^2 \theta_{23}=1/2$. These results confirm the analytical 
formulae in eq.(\ref{final2}); within this model it is difficult to produce large deviations of $\theta_{23}$ from maximal mixing  since large 
corrections can only be produced  invoking  
some fine-tuning among $\delta_{\nu_1}$ and the $y_i$'s.
  
\begin{figure}[h!]
\bc
\includegraphics[scale=.6]{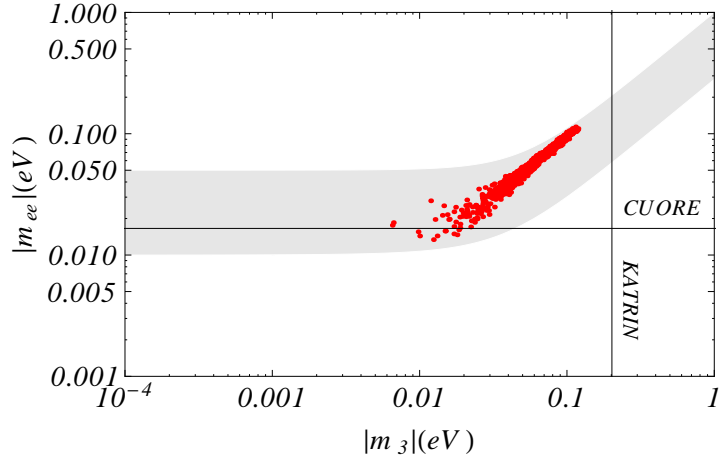}
\caption{\label{fig2}\it $|m_{ee}|$ as a function of the lightest mass $m_3$.
The shaded area corresponds to the region allowed by current neutrino data, for a mass ordering of inverted type.
The vertical band corresponds to the future sensitivity on the lightest neutrino mass of $0.2$ eV from 
the KATRIN experiment \cite{katrin} and the horizontal line to the future sensitivity of  $15$ meV  of CUORE \cite{cuore}.}
\ec
\end{figure}

In Fig.(\ref{fig2}) we plot the model prediction for $|m_{ee}|$ as a function of the absolute value of the lightest mass $m_3$.
The shaded area corresponds to the region allowed by current neutrino data, for a mass ordering of inverted type.
The vertical band corresponds to the future sensitivity on the lightest neutrino mass of $0.2$ eV from 
the KATRIN experiment \cite{katrin} and the horizontal line to the future sensitivity of $15$ meV  of CUORE \cite{cuore}.
The figure suggests that many of the model prediction points are in the quasi-degenerate region, above the sensitivity of CUORE. Furthermore, 
the values of the lightest neutrino mass rely in a region close to $|m_3|\sim \sqrt{\Delta m_{atm}^2}$, as it can be 
understood from eq.(\ref{masses}) and recalling that $v_u^2/\Lambda \sim \sqrt{\Delta m_{atm}^2}$.
As a consequence, much smaller values of $|m_3|$ can only be obtained if, thanks to additional fine tunings of the parameters,
a cancellation between the NLO and LO contributions takes place. Having obtained this only for very few points, we can 
consider the indication for a lower bound on $|m_3|$ around $5\cdot10^{-3}$ eV as a reasonable prediction of the model.

\subsection{The see-saw version} 
We introduce three right-handed neutrinos transforming under SU(5)$\times S_4\times Z_3$ as $(1,3_1,1)$ and carrying $U(1)_R=+1$.
At leading order the superpotential reads as follows:
\bea
\label{wd2}
 w_\nu &=& \frac{y_\nu}{\Lambda^{1/2}} F \nu^c H_5 +  M_\nu\Lambda\nu^c\nu^c + a (\nu^c\nu^c)_{3_1}  \varphi_\nu +  b (\nu^c\nu^c)_{1}  \xi_\nu 
\eea
and the resulting Dirac and Majorana mass matrices are:
\bea
m_\nu^D=s y v_u^0\left(
          \begin{array}{ccc}
            1 & 0 & 0 \\
            0 & 0 & 1 \\
            0 & 1 & 0 \\
          \end{array}
        \right)\qquad\qquad
M_N=\left(
              \begin{array}{ccc}
                M+b\lambda & -a\lambda & -a\lambda \\
                -a\lambda & 0 & M+b\lambda \\
                -a\lambda & M+b\lambda & 0 \\
              \end{array}
            \right)\Lambda\;,
\label{Feq:RHnu:masses}
\eea
similar to those quoted in \cite{Altarelli:2009gn}. In particular, the expression of the eigenvalues are the same and the spectrum results compatible with 
normal hierarchy only. We refer to that paper for a detailed discussion of the leading order results implied by eq.(\ref{Feq:RHnu:masses}) (and, in particular, of a proof 
that D=5 operators can be made subdominant).
At the next level of approximation, the additional contributions to $w_\nu$  are:
\bea
\label{nloneu}
\delta w_\nu &=& \frac{y_1}{\Lambda^{3/2}} (F \nu^c)_{3_1} H_5\varphi_\nu+ 
 \frac{y_2}{\Lambda^{3/2}} (F \nu^c)_1 H_5\xi_\nu \nn + \\ \nn \\  
&& +\frac{a_1}{\Lambda}(\nu^c\nu^c)(\varphi_\nu\varphi_\nu)+ \frac{a_2}{\Lambda}((\nu^c\nu^c)_2(\varphi_\nu\varphi_\nu)_2)+\\ \nn \\ 
   & &+ \frac{a_3}{\Lambda}((\nu^c\nu^c)_3(\varphi_\nu\varphi_\nu)_3)+ \frac{a_4}{\Lambda}(\nu^c\nu^c\varphi_\nu)\xi_\nu+h.c.\;,\nn
 \eea
Since there is no mixing between the flavon fields of different sectors, these contributions only modify the expressions of 
the neutrino masses but not the mixing matrix, which is still of BM form. However, as shown in eq.(\ref{vev:charged:best2}), 
the flavon $\varphi_\nu$ takes a non-vanishing component in the first element, which is the source of the BM breaking.
This affects the third operator in eq.(\ref{wd2}) and the first operator in eq.(\ref{nloneu}) and it is of the same order of magnitude
in  $\lambda$ of the contributions of the operators in eq.(\ref{nloneu}) with coefficients $a_1,..,a_4$.
To be more explicit and put in evidence the sources of the BM breaking, we write the leading deviation of the light neutrino mass matrix in the following way:
\bea
m_\nu=s^2 \left(v_u^{0}\right)^2\left(
          \begin{array}{ccc}
            x & y & y \\
            y & z+A & x-z \\
            y & x-z & z-A \\
          \end{array}
        \right),
\label{Feq:RHnu:masses2}
\eea
where
\bea
A=\frac{y_\nu}{M_\nu^2\Lambda}\, (2 M_\nu y_1 - a y_\nu)\,\delta_{\nu_1}\,\lambda^2\,.
\eea
and $x,y,z$ are linear combinations of the superpotential parameters. It is clear that the limit $A\to 0$ reproduces a mass matrix diagonalized by BM.
A straightforward calculation of the eigenvectors up to ${\cal O}(\lambda^2)$ shows that the solar and reactor angles do not receive any corrections 
whereas $\theta_{23}$ deviates from maximal mixing by a 
quantity proportional to $\delta_{\nu_1}\,\lambda^2$. 
Taking into account the corrections from the charged leptons, we arrive at a final result similar to eq.(\ref{final2}):
\bea
\label{final}
\theta_{13} &=&  \frac{1}{\sqrt{2} }\,\left(\frac{x_4^\prime}{x_5^\prime}\right)\,\lambda\nn \\ \nn \\
\theta_{23} &=&\frac{\pi}{4} + \left[-\frac{x_4^{\prime 2}}{4\, x_5^{\prime 2}}+\frac{x_6^2}{2}+x_7-\delta_{\nu_1} \left(\frac{a}{2 M_\nu}-\frac{y_1}{y_\nu}\right)\right]\lambda^2 \\ \nn \\
\theta_{12} &=&\frac{\pi}{4} -  \frac{1}{\sqrt{2}}\,\left(\frac{x_4^\prime +2\, x_5^\prime x_6 }{ \,x_5^\prime}\right)\,\lambda \nn.
\eea

\section{Conclusions}
\label{concl}
In this paper we have constructed an SU(5) model based on the flavour symmetry $S_4\times Z_3 \times U(1)_{FN}$, 
where the BM mixing is realized at the LO in a natural way. In order to get a realistic pattern of fermion masses and mixing and to get 
rid of the usual problems in SU(5) constructions, like the rigid mass relation $m_e=m_d^T$ and the doublet-triplet splitting problem,  
we have embedded the model in five dimensions,  with the fifth dimension compactified on the orbifold  $S^1/Z_2$.
In this way, we were able to reproduce the correct mass ratios in both charged lepton and quark sectors as well as a good agreement 
with the entries of the CKM mixing matrix, with only a moderate fine-tuning needed to describe the Cabibbo angle.
Since exact BM mixing is not confirmed by neutrino oscillation data,  large corrections are needed in order to lower the value of the solar angle.
Such corrections  arise at NLO through the diagonalization of the charged lepton mass matrix while, at the same time, the reactor angle receives corrections
of ${\cal O}(\lambda_C)$, which are perfectly in agreement with a large $\theta_{13}$ emerging from global fits to the world neutrino data. 
An  important feature of our model is that the shift of $\sin^2{\theta_{23}}$ from the maximal mixing value of 1/2 is 
expected to be of $\mathcal{O}(\lambda_C^2)$ at most. 
In order to reproduce the experimental value of the small parameter $r=\Delta m^2_{sol}/\Delta m^2_{atm}$ 
we need some amount of fine tuning. Here we choose to tune the phases of two complex numbers entering the neutrino mass matrix at leading order; 
this implies that the neutrino spectrum is mainly of the inverted hierarchy type (or moderately degenerate) and that  
the smallest light neutrino mass and the $0\nu \beta \beta$ parameter $|m_{ee}|$ are expected to be larger than about $0.01$ eV, as 
we obtained from a numerical analysis with mass matrices at the NLO.

\section*{Acknowledgments}
We thank Guido Altarelli and Luca Merlo for some interesting comments and discussions and Claudia Hagedorn for 
carefully reading the manuscript. We also acknowledge  MIUR (Italy), for financial support
under the contract PRIN08, and the CERN Theory Division where this work has been completed. 

\appendix
\section*{Appendix A: corrections to the leading order vacuum alignment}
\label{appvacuum}
In this appendix we explicitly compute the correction to the vacuum alignment of eq.(\ref{vev:neutrinos}).
For the sake of completeness, we use the following parametrization of the flavon vevs:
\bea
\langle\varphi_\ell\rangle &=&\nn v_{\varphi_\ell} \left(
                     \begin{array}{c}
                       \delta_{\varphi_1} \\
                       1+\delta_{\varphi_2}\\
                       \delta_{\varphi_3} \\
                     \end{array}
                   \right) \qquad\qquad
                    \langle\chi_\ell\rangle=v_{\chi }\left(
                     \begin{array}{c}
                       \delta_{\chi_1} \\
                       \delta_{\chi_2} \\
                       1\\
                     \end{array}
                   \right) \\
\langle\varphi_\nu\rangle& =&v_{\varphi_\nu} \left(
                     \begin{array}{c}
                        \delta_{\nu_1} \\
                       1+\delta_{\nu_2}\\
                       -1+ \delta_{\nu_3}\\
                     \end{array}
                   \right)  \qquad \langle\xi_\nu\rangle =v_{\xi} (1 + \delta v_\xi)\,.
\nn
\eea
The superpotential responsible for the vev shifts reads as follows:
\bea
\Lambda \Delta w_{d}&=& \varphi_\nu^0\left[\alpha_1(\varphi_\nu^2)_1 \varphi_\nu+ \alpha_2(\varphi_\nu^2)_2\varphi_\nu+ \alpha_3(\varphi_\nu^2)_{3_1} \varphi_\nu+
\alpha_4 (\varphi_\ell^2)_1 \varphi_\ell +\alpha_5 (\varphi_\ell^2)_2 \varphi_\ell+\alpha_6(\varphi_\ell^2)_{3_1} \varphi_\ell + \nn \right.\\ \nn&& \\
&&\left.\alpha_7(\chi_\ell^2)_2 \chi_\ell+\alpha_8 (\chi_\ell^2)_{3_1} \chi_\ell+ \alpha_9\xi_\nu (\varphi_\nu^2)_{3_1}+\alpha_{10}(\varphi_\ell^2)_2 \chi_\ell+
\alpha_{11}(\varphi_\ell^2)_{3_1} \chi_\ell+ \nn\right. \\ \nn&& \\ && \left. \alpha_{12}(\chi_\ell^2)_{1} \varphi_\ell+\alpha_{13}(\chi_\ell^2)_2 \varphi_\ell+
\alpha_{14}(\chi_\ell^2)_{3_1} \varphi_\ell+
\alpha_{15}\xi_\nu^2 \varphi_\nu\right]\nn + \\ \nn \\
&&\xi_\nu^0 \left[\beta_1(\varphi_\nu^2)_{3_1} \varphi_\nu+\beta_2\xi_\nu^3+\beta_3(\varphi_\ell^2)_{3_1} \varphi_\ell+
\beta_4\xi_\nu (\varphi_\nu^2)_{1}+\beta_5(\chi_\ell^2)_{3_1} \varphi_\ell\right] +\label{vevcorr} \\ \nn \\
&&\psi_\ell^0\left[
\gamma_1(\varphi_\ell^2)_{3_1} \varphi_\nu+\gamma_2
\xi_\nu  (\varphi_\ell^2)_{2} +\gamma_3(\chi_\ell^2)_{3_1} \varphi_\nu+\gamma_4
\xi_\nu(\chi_\ell^2)_{2}\right.+\nn \\ \nn \\ &&\left.\gamma_5
(\varphi_\nu \varphi_\ell)_{3_1} \chi_\ell+\gamma_6(\varphi_\nu \varphi_\ell)_{3_2} \chi_\ell+\gamma_7\xi_\nu(\varphi_\ell\chi_\ell)_{2}\right]\nn + \\ \nn \\
&&\chi_\ell^0\left[
\delta_1(\varphi_\ell^2)_{2} \varphi_\nu+\delta_2(\varphi_\ell^2)_{3_1} \varphi_\nu+\delta_3
(\chi_\ell^2)_{2}\varphi_\nu +\delta_4(\chi_\ell^2)_{3_1} \varphi_\nu+\right.\nn \\ \nn \\ 
&&\left.
\delta_5(\varphi_\nu \varphi_\ell)_{1} \chi_\ell+\delta_6(\varphi_\nu \varphi_\ell)_{2} \chi_\ell+\delta_7
(\varphi_\nu \varphi_\ell)_{3_1} \chi_\ell+\delta_8(\varphi_\nu \varphi_\ell)_{3_2} \chi_\ell+\delta_9\xi_\nu(\varphi_\ell\chi_\ell)_{3_2}
\right] \nn \,.
\eea
Even after considering the leading order relations in eq.(\ref{conditions}), some of the minimizing equations are still cumbersome. In the following we concentrate 
on those which easily allow to extract the relevant information on the vev shifts: 
\bea 
\frac{\partial \Delta w_d}{\partial \varphi^0_{\nu_1}} &=& 2 g_1 v_{\varphi_\nu}^2   (\delta_{\nu_2}+ \delta_{\nu_3}) = 0 \\ \nn \\
\label{eq2}\frac{\partial \Delta w_d}{\partial \psi^0_{\ell_1}} &=&   v_\chi \left[\sqrt{3} f_3 v_{\varphi_\ell} ( \delta_{\varphi_3}+\delta_{\chi_2})-
2 f_2 \delta_{\chi_2}  v_\chi\right]-2  f_1 \delta_{\varphi_3}  v_{\varphi_\ell}^2 \\ \nn \\
\frac{\partial \Delta w_d}{\partial \varphi^0_{\chi_1}} &=&  4  v_{\varphi_\ell}  v_\chi  (\delta_{\varphi_3}- \delta_{\chi_2}) = 0 \,.
\eea
We see that the last equation implies $\delta_{\varphi_3}= \delta_{\chi_2}$ which, using eq.(\ref{eq2}), forces $\delta_{\varphi_3}= \delta_{\chi_2}=0$.
The first equation tells us that $\delta_{\nu_2}=- \delta_{\nu_3}$ and that they are undetermined.
Using the previous relations, the remaining minimizing equations allow to fix the magnitude of the other vev shifts in terms of the superpotential parameters. 
In particular, from $ \partial \Delta w_d/\partial \varphi^0_{\nu_2}$ and  $\partial \Delta w_d/ \partial \varphi^0_{\nu_3}$ we get 
a set of conditions on $\delta_{\nu_1}$ and $\delta v_\xi$, which result both different from zero. From 
$ \partial \Delta w_d/\partial \chi^0_{2,3} = 0$ we obtain an expression for $\delta_{\chi_1}$ and $\delta_{\varphi_1} $, respectively; the last
equation  $ \partial \Delta w_d/\psi^0_{\ell_2} = 0$ depends on  $\delta_{\varphi_2}$, which  is also completely determined by the superpotential parameters. 
\section*{Appendix B: corrections to the fermion mass matrices}
In this appendix we quote all the operators with three-flavon insertions responsible for the down quark sector:
\bea
F T_3 H_{\bar 5}&:&\qquad  (\varphi_\nu^2)_1 \varphi_\ell,\; (\varphi_\nu^2)_2 \varphi_\ell,\; 
(\varphi_\nu^2)_{3_1} \varphi_\ell,\; (\varphi_\nu^2)_{2} \chi_\ell \nn \\ \nn \\
&& \qquad(\varphi_\nu^2)_{3_1}\chi_\ell ,\; \xi_\nu^2 \varphi_\ell,\; (\varphi_\nu\varphi_\ell)_{3_1} \xi_\nu,\; 
(\varphi_\nu\chi_\ell)_{3_1} \xi_\nu
\nn \\ \nn \\
F T_2 H_{\bar 5}\theta&:&\qquad (\varphi_\nu^2)_1 \varphi_\nu,\; (\varphi_\nu^2)_2 \varphi_\nu,\; 
(\varphi_\nu^2)_{3_1} \varphi_\nu,\; (\varphi_\ell^2)_{1} \varphi_\ell \nn \\ \nn \\
&& \qquad(\varphi_\ell^2)_2 \varphi_\ell,\; (\varphi_\ell^2)_{3_1} \varphi_\ell,\; (\chi_\ell^2)_{2} \chi_\ell,\; 
(\chi_\ell^2)_{3_1} \chi_\ell,\;(\varphi_\nu^2)_{3_1} \xi_\nu,\;\xi_\nu^2\varphi_\nu,\; \nn \\ \nn \\
&&\qquad
(\varphi_\ell^2)_{2} \chi_\ell ,\;(\varphi_\ell^2)_{3_1} \chi_\ell,\;(\chi_\ell^2)_{1} \varphi_\ell
,\;(\chi_\ell^2)_{2} \varphi_\ell,\;(\chi_\ell^2)_{3_1} \varphi_\ell
\nn \\ \nn \\
F T_2 H_{\bar 5}\theta^\prime&:&\qquad  (\varphi_\ell^2)_1 \varphi_\nu,\; (\varphi_\ell^2)_2 \varphi_\nu,\;(\varphi_\ell^2)_{3_1} \varphi_\nu,\;
(\varphi_\ell^2)_{3_1} \xi_\nu,\; (\chi_\ell^2)_1 \varphi_\nu,\;(\chi_\ell^2)_{2} \varphi_\nu,\; \nn \\ \nn \\
&& \qquad(\chi_\ell^2)_{3_1} \varphi_\nu,\; (\chi_\ell^2)_{3_1} \xi_\nu,\; (\varphi_\nu\varphi_\ell)_{2}\chi_\ell,\;
(\varphi_\nu\varphi_\ell)_{3_1} \chi_\ell, \nn \\   \\
&& \qquad(\varphi_\ell\varphi_\nu)_{3_2}\chi_\ell,\;(\varphi_\ell\chi_\ell)_{3_1}\xi_\nu\nn \\ \nn \\
F T_1 H_{\bar 5}\theta^3&:&\qquad  \text {the same as $F T_2 H_{\bar 5}\theta^\prime$} \nn\\ \nn \\
F T_1 H_{\bar 5}\theta^2\theta^\prime&:&\qquad \text {the same as $F T_3 H_{\bar 5}$} \nn\\ \nn \\
F T_1 H_{\bar 5}\theta\theta^{\prime 2}&:&\qquad \text {the same as $F T_2 H_{\bar 5}\theta$} \nn\\ \nn \\
F T_1 H_{\bar 5}\theta^{\prime 3}&:&\qquad \text{the same as $F T_2 H_{\bar 5}\theta^\prime$} \nn
\eea

\section*{Appendix C: the group $S_4$}
We report here the multiplication table for $S_4$ and we list the Clebsch-Gordan coefficients in the basis used in the paper \cite{Altarelli:2009gn}. In the following we
use $\alpha_i$ to indicate the elements of the first representation of the product and $\beta_i$ to indicate those of the second representation.
Explicit forms of $S$ and $T$ are as follows. In the representation $1$ we have $T=1$ and $S=1$, while $T=-1$ and $S=-1$ in $1'$. In the representation $2$ we have:
\bea
T=\left(
    \begin{array}{cc}
      1 & 0 \\
      0 & -1 \\
    \end{array}
  \right)\qquad\qquad
S=\frac{1}{2}\left(
    \begin{array}{cc}
      -1 & \sqrt3  \\
      \sqrt3 & 1   \\
    \end{array}
  \right)
  \label{ST2}\;.
\eea
For the representation $3$, the generators are:
\bea
T=\left(
    \begin{array}{ccc}
      -1 & 0 & 0 \\
      0 & -i & 0 \\
      0 & 0 & i \\
    \end{array}
  \right)\qquad\qquad
S=\left(
    \begin{array}{ccc}
      0 & -\frac{1}{\sqrt{2}} & -\frac{1}{\sqrt{2}}  \\
      -\frac{1}{\sqrt{2}} & \frac{1}{2} & -\frac{1}{2} \\
      -\frac{1}{\sqrt{2}} & -\frac{1}{2} & \frac{1}{2} \\
    \end{array}
  \right)\;.
  \label{matS}
\eea
In the representation $3'$ the generators $S$ and $T$ are simply opposite in sign with respect to those in the 3.

We start with all the multiplication rules which include the 1-dimensional representations:
\[
\begin{array}{l}
1\otimes Rep=Rep\otimes1=Rep\quad\rm{with~Rep~any
~representation}\\[-10pt]
\\[8pt]
1_2\otimes1_2=1\sim\alpha\beta\\[-10pt]
\\[8pt]
1_2\otimes2=2\sim\left(\begin{array}{c}
                    \alpha\beta_2 \\
                    -\alpha\beta_1 \\
            \end{array}\right)\\[-10pt]
\\[8pt]
1_2\otimes3_1=3_2\sim\left(\begin{array}{c}
                    \alpha\beta_1 \\
                    \alpha\beta_2 \\
                    \alpha\beta_3\\
                    \end{array}\right)\\[-10pt]
\\[8pt]
1_2\otimes3_2=3_1\sim\left(\begin{array}{c}
                            \alpha\beta_1 \\
                            \alpha\beta_2 \\
                            \alpha\beta_3\\
                    \end{array}\right)
\end{array}
\]
The multiplication rules with the 2-dimensional
representation are the following ones:
\[
\begin{array}{ll}
2\otimes2=1\oplus1_2\oplus2&\quad
\rm{with}\quad\left\{\begin{array}{l}
                    1\sim\alpha_1\beta_1+\alpha_2\beta_2\\[-10pt]
                    \\[8pt]
                    1_2\sim\alpha_1\beta_2-\alpha_2\beta_1\\[-10pt]
                    \\[8pt]
                    2\sim\left(\begin{array}{c}
                        \alpha_2\beta_2-\alpha_1\beta_1 \\
                        \alpha_1\beta_2+\alpha_2\beta_1\\
                    \end{array}\right)
                    \end{array}
            \right.\\[-10pt]
\\[5pt]
2\otimes3_1=3_1\oplus3_2&\quad
\rm{with}\quad\left\{\begin{array}{l}
                    3_1\sim\left(\begin{array}{c}
                        \alpha_1\beta_1\\
                        \frac{\sqrt3}{2}\alpha_2\beta_3-\frac{1}{2}\alpha_1\beta_2 \\
                        \frac{\sqrt3}{2}\alpha_2\beta_2-\frac{1}{2}\alpha_1\beta_3 \\
                    \end{array}\right)\\[-10pt]
                    \\[8pt]
                    3_2\sim\left(\begin{array}{c}
                        -\alpha_2\beta_1\\
                        \frac{\sqrt3}{2}\alpha_1\beta_3+\frac{1}{2}\alpha_2\beta_2 \\
                        \frac{\sqrt3}{2}\alpha_1\beta_2+\frac{1}{2}\alpha_2\beta_3 \\
                    \end{array}\right)\\
                    \end{array}
            \right.\\[-10pt]
\\[5pt]
2\otimes3_2=3_1\oplus3_2&\quad
\rm{with}\quad\left\{\begin{array}{l}
                    3_1\sim\left(\begin{array}{c}
                       -\alpha_2\beta_1\\
                        \frac{\sqrt3}{2}\alpha_1\beta_3+\frac{1}{2}\alpha_2\beta_2 \\
                        \frac{\sqrt3}{2}\alpha_1\beta_2+\frac{1}{2}\alpha_2\beta_3 \\
                    \end{array}\right)\\[-10pt]
                    \\[8pt]
                    3_2\sim\left(\begin{array}{c}
                        \alpha_1\beta_1\\
                        \frac{\sqrt3}{2}\alpha_2\beta_3-\frac{1}{2}\alpha_1\beta_2 \\
                        \frac{\sqrt3}{2}\alpha_2\beta_2-\frac{1}{2}\alpha_1\beta_3 \\
                    \end{array}\right)\\
                    \end{array}
            \right.\\
\end{array}
\]

The multiplication rules involving the 3-dimensional
representations are:
\[
\begin{array}{ll}
3_1\otimes3_1=3_2\otimes3_2=1\oplus2\oplus3_1\oplus3_2\qquad
\rm{with}\quad\left\{
\begin{array}{l}
1\sim\alpha_1\beta_1+\alpha_2\beta_3+\alpha_3\beta_2\\[-10pt]
                    \\[8pt]
2\sim\left(
     \begin{array}{c}
       \alpha_1\beta_1-\frac{1}{2}(\alpha_2\beta_3+\alpha_3\beta_2)\\
       \frac{\sqrt3}{2}(\alpha_2\beta_2+\alpha_3\beta_3)\\
     \end{array}
   \right)\\[-10pt]
   \\[8pt]
3_1\sim\left(\begin{array}{c}
         \alpha_3\beta_3-\alpha_2\beta_2\\
         \alpha_1\beta_3+\alpha_3\beta_1\\
         -\alpha_1\beta_2-\alpha_2\beta_1\\
        \end{array}\right)\\[-10pt]
        \\[8pt]
3_2\sim\left(\begin{array}{c}
         \alpha_3\beta_2-\alpha_2\beta_3\\
         \alpha_2\beta_1-\alpha_1\beta_2\\
         \alpha_1\beta_3-\alpha_3\beta_1\\
    \end{array}\right)
\end{array}\right.\\\\[10pt]
3_1\otimes3_2=1_2\oplus2\oplus3_1\oplus3_2\qquad
\rm{with}\quad\left\{
\begin{array}{l}
1_2\sim\alpha_1\beta_1+\alpha_2\beta_3+\alpha_3\beta_2\\[-10pt]
        \\[8pt]
2\sim\left(
     \begin{array}{c}
     \frac{\sqrt3}{2}(\alpha_2\beta_2+\alpha_3\beta_3)\\
     -\alpha_1\beta_1+\frac{1}{2}(\alpha_2\beta_3+\alpha_3\beta_2)\\
     \end{array}
   \right)\\[-10pt]
        \\[8pt]
3_1\sim\left(\begin{array}{c}
         \alpha_3\beta_2-\alpha_2\beta_3\\
         \alpha_2\beta_1-\alpha_1\beta_2\\
         \alpha_1\beta_3-\alpha_3\beta_1\\
    \end{array}\right)\\[-10pt]
        \\[8pt]
3_2\sim\left(\begin{array}{c}
         \alpha_3\beta_3-\alpha_2\beta_2\\
         \alpha_1\beta_3+\alpha_3\beta_1\\
         -\alpha_1\beta_2-\alpha_2\beta_1\\
    \end{array}\right)\\
\end{array}\right.
\end{array}
\]

\end{document}